\documentclass[unpublished,letterpaper]{quantumarticle}
\pdfoutput=1
\usepackage[utf8]{inputenc}
\usepackage[english]{babel}
\usepackage{amsmath}
\usepackage{amssymb,bm}
\usepackage{mathrsfs}
\usepackage{amsfonts}
\usepackage{bm}
\usepackage{graphicx}
\usepackage{endnotes}
\usepackage{array}
\usepackage{url}
\usepackage{enumitem}
\usepackage{quoting}
\newcommand{\ket}[1]{|#1\rangle}

\usepackage[
    hyperfootnotes=true         
    ]{hyperref}                 

\hypersetup{
colorlinks=true,                
linkcolor=red,                  
anchorcolor=black,              
citecolor=blue,                
urlcolor=cyan,                  
bookmarksnumbered=true,         
linktocpage=false               
}
\begin{document}
%
%

%
%
%

\title{The Hamiltonian for Entangled States Cannot Be Additive}

\author{Kent A. Peacock}

\affiliation{Department of Philosophy, University of Lethbridge\\ 4401 University Drive, Lethbridge, Alberta, Canada.  T1K 3M4\\  \url{kent.peacock@uleth.ca}  }

\begin{abstract}
The assumption that the system Hamiltonian for entangled states is additive is widely used in orthodox quantum no-signalling arguments.  It is shown that additivity implies a contradiction with the assumption that the system being studied is entangled.
\end{abstract}


\maketitle

\section{Introduction:   Is there a `Bell Telephone'?}
A large literature exists on the question of whether it is possible to use quantum nonlocality to signal superluminally in a controllable fashion.  The conventional wisdom is that this is out of the question in quantum mechanics as it is currently formulated (e.g., \cite{GRW1980,Dieks1982,Shimony1983, Jordan1983,Redhead1987,EbRoss1989}), although a dissenting literature exists   \cite{Bussey1987,Peacock1991b,Kennedy1995,Mitt1998,Peacock2018}.  

Many no-controllable signalling (NCS) arguments depend, directly or indirectly, upon the assumption that the Hamiltonian for entangled states is of a local or additive form.  
The aim of this note is to show that this assumption can be undercut by a very simple \emph{reductio ad absurdum} argument.  

An example of a NCS argument that uses the additivity assumption appears a widely-cited paper by A. Shimony \cite{Shimony1983}. Shimony considers a two-particle system (which presumably could be entangled) and \emph{assumes without argument} that if we label the  particles $A$ and $B$, the Hamiltonian for the system of two particles (not including interactions it may have with measuring devices) has the form
\begin{equation}
  H_{AB} = H_A \otimes \mathbb{I}_B + \mathbb{I}_A \otimes H_B,
  \label{ShimHam}
\end{equation}
where  $\mathbb{I}_B$ is identity in $B$'s subspace, and correspondingly for $A$.  Shimony also assumes that the measurement procedure  applied to one particle (say $A$)  acts only locally on $A$.  Under these strong locality assumptions Shimony shows that the time evolution operator associated with the measurement process, $U(t) = \exp{(-H_{AB}t/\hbar)}$, acts only trivially on the other particle;  nothing done to the first particle can affect the local statistics measureable on the second.  It is well known that changing the relative detector parameter in entangled states affects the \emph{correlation} between results taken on $A$ and $B$; thus one can encode a message in the correlations, a fact which is the basis for quantum cryptography.  But this can be read only by having both local sequences of results, and thus it does not constitute a method of controllable nonlocal signalling  as such.  

An interesting feature of Shimony's argument is that it does not use the collapse hypothesis; rather, the argument takes it that a measurement acting on one particle will evolve the system in a unitary fashion.  The question of signalling does not turn on whether the wave function collapses, therefore, but on getting the correct description for the dynamics of the system consisting of an apparatus coupled to an entangled multiparticle state.

An approach similar to Shimony's is taken by M.\ Redhead \cite[\S 4.6]{Redhead1987}, and D.\ Dieks also presents a NCS argument that depends upon the assumption that a Hamiltonian can be assigned to the individual particles in an entangled state \cite{Dieks1982}.  The question now is whether the additivity assumption is justified.  

\section{A \textit{Reductio} of the Additivity Assumption}  
We show here that there is a straightforward \textit{reductio ad absurdum} argument against the additivity assumption for entangled states.  
To begin with, we cite two facts about the mathematics of entangled states.  

Rule 1: It is well known that subsystems of a tensor product (entangled) state cannot be pure states.  As Cohen-Tannoudji \textit{et al.} put it,
\begin{quote}
 "[A]n interaction between the two systems transforms an initial state which is a product into one which is no longer a product:  any interaction between two systems therefore introduces, in general, correlations between them.  \dots  This question is very important since, in general,  every physical system has interacted with others in the past \dots  \emph{it is not possible to associate a state vector  $\ket{\phi(1)}$   [a pure state]  with system (1) alone}.'' [emphasis added]  \cite[p.~293]{CT}.
\end{quote}
Subsystems of a tensor product state must be described as mixed states---classical probability distributions over sets of pure states.  

Rule 2:  To say that an observable is associated with a state $\ket{\psi}$ is to say that the observable has a set of eigenstates which define a basis for the state space in which $\ket{\psi}$ lives \cite[Chap II.D]{CT}.

Now suppose that it is possible to associate a Hamiltonian $H_1$ with a particle
$p_1$ belonging to an entangled multiparticle system.  By Rule 2,  it would be possible to write the state of
that particle as an expansion of the form
\begin{equation}
  \ket{\phi(p_1)} = \sum_i c_i\ket{e_i}  \label{A}
\end{equation}
where $\{\ket{e_i} \}$ are the energy eigenstates of the particle $p_1$ with respect to $H_1$, and $c_i$ are complex coefficients.  \emph{Any expression of this form is a pure state}, since it is a linear combination of pure states (the presumed local energy eigenstates of $H_1$).  
 But by  Rule 1, because $p_1$ is taken to be a member of an entangled state, it cannot, by itself, itself be represented as a pure state---otherwise, it would not be entangled.  Hence, there cannot exist a Hamiltonian that can be associated with $p_1$ in this way.  

In sum, the assumption of additivity \eqref{ShimHam} contradicts the assumption that the particles are entangled.  
Thus, any no-signalling argument that depends upon the assumption of additivity (or any other equivalent expression of dynamic localizability) is merely a demonstration of no-signalling for \emph{product states}, which \emph{do} have additive dynamics.

Just as the state of a particle in an entangled state is a mixture (a classical probability distribution over possible pure states), the possible energies for individual particles in entangled states must be a classical probability distribution over possible energy states.  As Cohen-Tannoudji \textit{et al.} observe,
\begin{quote}
``Just as with vectors, there exist operators in $\mathscr{E}$ [a tensor product space] which are not tensor products of an operator in $\mathscr{E}_1$ and an operator of $\mathscr{E}_2$." \cite[p.\ 157]{CT}
\end{quote}
The system Hamiltonian for an entangled tensor product space belonging to a multiparticle system must be such a global or nonlocal operator.  

The energy of one particle in an entangled system must in general, therefore, depend nonlocally on what is done to the other remote particles in the system.  As Shimony himself correctly emphasized \cite{Shimony1983}, whether or not this dependency is locally controllable is an important but distinct question.  
However, the \textit{reductio} argument outlined here rules out any NCS argument depending upon a prior assumption of the dynamic locality of the subsystem particles.     

\section{Dynamics of Entangled States in Quantum Information Theory}
What, then, is the correct description of the dynamics of entangled states?  In 1933, W.\ Pauli stated, 
\begin{quote}
``An additive decomposition of the Hamiltonian into independent summands corresponds [\emph{entsprich}], therefore, to a product decomposition of the wavefunction into independent factors.'' \cite{Pauli33,Howard_Talk}
\end{quote}
 Unfortunately, Pauli did not directly address the question of what would pertain to the dynamics of \emph{nonfactorizable} multiparticle states and so it is unclear, from his exposition alone, whether his \textit{entsprich} should be read as ``if and only if".  

In fact, it is a commonplace in recent literature on quantum information theory that the Hamiltonians for entangled states themselves contain cross-terms and thus cannot in general be expressed in additive form.  For example,  D{\"u}r \textit{et al.}  (2002) give the general Hamiltonian for an entangled (Bell) state of two particles as,
\begin{equation}
  H_{AB} = \sum_{i=1}^3 \alpha_i\sigma_i^A \otimes \mathbb{I}_B + \sum_{j=1}^3\beta_j \mathbb{I}_A \otimes \sigma_j^B + \sum_{i,j=1}^3\gamma_{i,j}\sigma_i^A \otimes \sigma_j^B,
\label{Thing} 
\end{equation}
where the $\sigma_i$ are the Pauli matrices.  
 This expansion contains irreducible cross-terms as shown, and therefore cannot be represented in purely additive form.  Thus, again, we see that any NCS argument that depends upon Shimony's additivity assumption \eqref{ShimHam} cannot apply without qualification to entangled states, which are the only sort of states of interest in the signalling question.

The cross-terms in \eqref{Thing} suggest that the full energy spectrum of entangled states must contain terms that are not local to the individual components of the system, but which (like the energies of electron orbitals in atoms) are properties of the system as a whole.  
An interferometric experiment by Lee \textit{et al.} \cite{LeeKC11,Grossman11} directly demonstrates the existence of nonlocal energy states in certain kinds of entangled systems. In this experiment, two diamond chips 30 cm apart are demonstrably put into the same phonon state. This experimental finding alone indicates that the additivity assumption for entangled states cannot be generally correct.

\section{Conclusion and Discussion}

In sum:  if the Hamiltonian of a multiparticle system is additive (entirely local to the particles), then the system cannot be entangled---since otherwise it would be impossible to assign a spectrum of pure states to the component particles as in \eqref{A}.  If this view is correct, then any NCS arguments that rely on the additivity assumption amount merely to unobjectionable demonstrations of NCS for product states.  This observation, by itself, does not show that controllable quantum signalling is possible, but it shows that a large class of arguments commonly cited against that possibility are simply inapplicable to the only kind of systems (entangled states) for which  the question of signalling arises.  

The other major route that has been used to argue for NCS is \textit{via} microcausality (the assumption that measurement operations on spacelike separate particles in a multiparticle system always commute) \cite{Mitt1998}.  The argument presented in this note does not fully address the question of microcausality.  We simply note the following points.  If operations on individual spacelike separate particles in a multiparticle system act only locally on their respective particles, then it is readily shown that they commute in their action on the system as a whole, as one would expect.  On the other hand, if  operators are global in the sense indicated by Cohen-Tannoudji \textit{et al.}, it is not immediately apparent that they can be guaranteed to commute in their action on an entangled state considered \emph{as a whole} \cite{Peacock1991b}.   This important question demands further study.  

\section{Acknowledgements}
The author thanks the University of Lethbridge and the Social Sciences and Humanities Research Council of Canada for support.  Thanks also to the following for helpful discussion:   J.R. Brown, M.B.\ Brown, S.\ Das,  B.\ Hepburn, and W.\ D{\"u}r.  None of these people or organizations are responsible for any errors in this paper.



\end{document}